\documentclass[aps,prc,twocolumn,floatfix,superscriptaddress]{revtex4}
\usepackage{longtable}
\usepackage{graphicx}
\usepackage{dcolumn}
\usepackage{color}
\usepackage{xcolor}
\usepackage{float}
\usepackage{amsmath}
\usepackage{soul}
\usepackage{multirow}
\usepackage{booktabs}
\usepackage{url}
\usepackage{comment}
\usepackage{slashed}

\begin{document}

\title{Hindered $\Delta K=1$ Dipole Strength in octupole bands in $N=90$  $^{154}$Gd from Lifetime Measurements with $\gamma-\gamma$ fast timing technique}

\author{A.~Pal}
\affiliation{Variable Energy Cyclotron Centre, Kolkata - 700 064, India}
\affiliation{Homi Bhabha National Institute, Training School Complex, Anushakti Nagar, Mumbai - 400 094, India}
\author{S.~Basak}
\thanks{Present address: Institute of Modern Physics, Chinese Academy of Sciences, Lanzhou, Gansu - 730 000, China}
\affiliation{Variable Energy Cyclotron Centre, Kolkata - 700 064, India}
\affiliation{Homi Bhabha National Institute, Training School Complex, Anushakti Nagar, Mumbai - 400 094, India}
\author{D.~Kumar}
\affiliation{Variable Energy Cyclotron Centre, Kolkata - 700 064, India}
\affiliation{Homi Bhabha National Institute, Training School Complex, Anushakti Nagar, Mumbai - 400 094, India}
\author{T.~Bhattacharjee}
\thanks{Corresponding author}
\email{btumpa@vecc.gov.in}
\affiliation{Variable Energy Cyclotron Centre, Kolkata - 700 064, India}
\affiliation{Homi Bhabha National Institute, Training School Complex, Anushakti Nagar, Mumbai - 400 094, India}
\author{B.~Maheshwari}
\affiliation{Variable Energy Cyclotron Centre, Kolkata - 700 064, India}
\author{K.~Nomura}
\affiliation{Department of Physics, Hokkaido University, Sapporo 060-0810, Japan} 
\affiliation{Nuclear Reaction Data Center, Hokkaido University, Sapporo 060-0810, Japan}
\author{P.~Van~Isacker}
\affiliation{Grand Acc{\`e}l{\` e}rateur National d'Ions Lourds, 14000 Caen, France}
\author{D.~Banerjee}
\affiliation{Variable Energy Cyclotron Centre, Kolkata - 700 064, India}
\affiliation{Homi Bhabha National Institute, Training School Complex, Anushakti Nagar, Mumbai - 400 094, India}
\author{S.~S.~Alam}
\affiliation{Government General Degree College, Chapra - 741 123, West Bengal, India}
\author{A.~K.~Jain}
\affiliation{Amity Institute of Nuclear Science and Technology, Amity University Noida 201313, India}

\date{\today}
\begin{abstract}
The lifetimes of the low-lying negative-parity $1^-$ state at 1414~keV and $2^-$ state at 1398~keV in $^{154}$Gd have been measured using the $\gamma$--$\gamma$ fast-timing technique with the VENTURE array
at VECC, Kolkata. The states were populated through the $\beta$ decay of $^{154}$Tb, produced in proton-induced reactions at the K130 cyclotron.
From the measured lifetimes, absolute $B(E1)$ transition strengths were deduced. The extracted $B(E1)$ values are compared with those of neighboring Gd isotopes and with Gogny-HFB-based $sdf$-IBM calculations. The results show that the measured $E1$ strengths from these states are strongly hindered compared with the corresponding $\Delta K=0$ transitions, providing
evidence for weak $\Delta K=1$ dipole strength in $^{154}$Gd.
\end{abstract}

\maketitle

\section{Introduction}
\label{intro}

Octupole correlations constitute an important mode of collectivity in atomic nuclei and are associated with reflection-asymmetric degrees of freedom of the nuclear surface~\cite{buttler1,stanislaw,ahmad}. In quadrupole-deformed systems, an octupole vibration built on the deformed shape gives rise to low-lying negative-parity rotational structures and  characteristic $E1$ transitions linking opposite-parity bands. In such a picture, the octupole and quadrupole degrees of freedom can generate an intrinsic dipole moment, making enhanced $E1$ strengths 
a widely used fingerprint of octupole correlations. Systematic investigations of low-lying $E1$ strength in rare-earth nuclei emphasize that absolute $B(E1)$ values provide a particularly sensitive discriminator of the microscopic origin of dipole excitations \cite{spieker2015}. The absolute $B(E1)$ values are especially informative because they are sensitive to detailed nuclear wave functions: even small admixtures can modify the dipole strength significantly \cite{cottle1996}.  

Octupole collectivity is firmly established in regions such as the
actinides and  the Ba–Sm region—where alternating-parity bands,
parity doublets, and enhanced $E1$ strengths are well documented \cite{fransen,peker}. However, the situation in heavier rare-earth nuclei,
including the Gd isotopes, remains less constrained experimentally. Very recently, 
an enhanced $B(E3;3^- \!\rightarrow\! 0^+)$ value has been reported in $^{150}$Gd, providing direct evidence for large octupole collectivity within the Gd isotopic chain \cite{pascu2025}. 

In deformed nuclei, the negative-parity octupole excitation can split into rotational bands characterized by the projection quantum number $K$, with low-lying $K^\pi=0^-$ and $K^\pi=1^-$ structures often being the most prominent ones~\cite{greiner}. When $K$ is approximately conserved, $\Delta K=1$ $E1$ transitions are typically hindered relative to $\Delta K=0$ transitions; and the relative $\Delta K=0$ and $\Delta K=1$ strengths offer a sensitive probe of $K$-mixing and Coriolis coupling \cite{Kocbach1970,konjin}. From a qualitative microscopic perspective, Börner \textit{et al.}~\cite{borner} explained this hindrance using a Nilsson-based counting argument, in which the number of available $\Delta K=0$ $E1$ configurations near the Fermi surface is larger than that of the $\Delta K=1$ configurations, with the latter decreasing more rapidly with increasing neutron number. 
Coulomb-excitation studies of $^{154}$Gd by Sugawara et al.~\cite{sugawara} reported a sizable dipole strength associated with the $K^\pi=0^-$ band, whereas no comparable enhancement was observed for the $K^\pi=1^-$ structure. Therefore, absolute $B(E1)$ values are essential for quantifying the degree of $\Delta K=1$ hindrance and may also be used for K assignments to the negative parity levels indirectly.

The even-even nucleus $^{154}$Gd ($Z=64$, $N=90$) occupies a key position in the Gd isotopic chain in the transitional region around $N = 90$ ~\cite{castenE4}. It exhibits several low-lying negative-parity levels commonly discussed in terms of octupole-vibrational sequences with intrinsic projections $K^\pi=0^-$, $K^\pi=1^-$, and $K^\pi=2^-$ \cite{vogel,west}. However, the $K^\pi$  assignment of the lowest negative-parity states in $^{154}$Gd has a somewhat complex history. Based largely on $\log ft$ arguments, Meyer \textit{et al.}~\cite{meyer}
discussed the low-spin negative-parity excitations of $^{154}$Gd as $K^\pi=0^-$, $1^-$, and $2^-$ octupole bands. The $K^\pi=0^-$ sequence proposed in this work consisted of the lowest odd-spin negative parity states and the $K^\pi=1^-$ sequence showed the unusual low-spin pattern $I^\pi=2^-,1^-,4^-,3^-$ rather than the conventional $1^-,2^-,3^-,4^-$ ordering. This inversion was interpreted as arising from strong rotation--vibration (Coriolis) coupling between the $K^\pi=1^-$ and $K^\pi=0^-$ octupole bands. Subsequent theoretical works also adopted the above $K$ assignments~\cite{vogel,casten1993}. Casten \textit{et al.} \cite{casten1993} showed that the relative ordering of the $K^\pi=0^-$ and $K^\pi=1^-$ octupole bands evolves rapidly with neutron number in the Gd isotopes; for example, the sequence of the $K^\pi=0^-$ and $K^\pi=1^-$ bands differs between $^{154}$Gd and $^{156}$Gd.  In the early $\beta$-decay studies of $^{154}$Eu \cite{kulp2003,kulp2004} and the $\varepsilon$ decay of $^{154}$Tb \cite{sousa1975} also followed the same assignment and the yrast negative-parity states were interpreted as members of the $K^\pi=0^-$ octupole-vibrational band. A similar interpretation was adopted in the in-beam study of Morrison \textit{et al.} \cite{morrison} using the $^{150}$Nd($^{9}$Be,5n$\gamma$) reaction, where the higher-spin negative parity yrast levels were also associated with the $K^\pi=0^-$ structure.  In contrast to all the above studies, only recently, Sharpey-Schafer \textit{et al.} \cite{sharpey}, based on $\gamma$-ray spectroscopy using the reactions
$^{152}$Sm($\alpha$,2n)$^{154}$Gd and
$^{154}$Sm($\alpha$,4n)$^{154}$Gd with the AFRODITE array, showed the lowest odd spin negative-parity states as members of the $K^\pi=1^-$ band instead of $K^\pi=0^-$. This assignment was subsequently adopted in the ENSDF adopted levels~\cite{ensdf,nndc}; however, in contrast, the ENSDF decay datasets still continue to associate the yrast low-lying negative-parity states with the $K^\pi=0^-$ octupole-vibrational band.
In this context, the $B(E1)$ values from the low lying negative parity levels becomes a particularly important experimental observable for assessing their band character. For the heavier Gd isotopes, however, lifetimes are predominantly known only for low-lying states of the $K^\pi=0^-$ band, whereas corresponding data for $K^\pi=1^-$ members are limited; in particular, the lowest even-spin $2^-$ state has not been measured directly in neighboring isotopes and only upper limits are available. 

In the present work, we report the first lifetime measurements of the two low-lying negative-parity levels in $^{154}$Gd, namely the $2^-$ state at $E_x=1397.6$~keV and the $1^-$ state at $E_x=1414$~keV  while the lifetime for the lowest $1^-$ state at 1241~keV is known\cite{ensdf}. The states were populated following the $\beta$ decay of $^{154}$Tb produced with proton-induced reaction at the K130 cyclotron; and the lifetimes were determined using the $\gamma$--$\gamma$ fast-timing technique with the VENTURE array \cite{venture}. The extracted lifetimes yield absolute $B(E1)$ values for the decay of these states enabling their quantitative comparison and, therefore, the $K$ assignments. The results are compared with $sdf$-IBM calculations and interpreted with a band mixing calculation to understand the role of configuration mixing in the hindrance of $\Delta K=1$ $E1$ strength.

\section{Experimental Details}
\label{exp}

The low-lying excited states of $^{154}$Gd were populated following the radioactive decay of the three
long-lived isomers of $^{154}$Tb ($I^\pi=3^-$, $t_{1/2}=9.4$~h; $I^\pi=7^-$, $t_{1/2}=22.7$~h; and
$I^\pi=0$, $t_{1/2}=21.5$~h). The $^{154}$Tb activity was produced via two complementary reactions,
$^{154}$Gd(p,n)$^{154}$Tb and $^{\text{nat}}$Eu($\alpha$,4n)$^{154}$Tb, using proton and $\alpha$ beams of
12~MeV and 40~MeV, respectively, delivered by the K--130 cyclotron at VECC, Kolkata.
For the present lifetime measurement, the proton-induced reaction on an enriched $^{154}$Gd target
(67\% enrichment) was preferred, as it yields a substantially cleaner $\gamma$--$\gamma$ coincidence
environment compared to the $\alpha$-induced reaction on natural Eu. The latter tends to populate higher-spin
isomeric components in $^{154}$Tb and also produces additional contaminant activities (notably from $^{156}$Tb)
due to the isotopic composition of the natural Eu target. Accordingly, the results reported here are based
on the dataset acquired following the $^{154}$Gd(p,n)$^{154}$Tb reaction.

The $\beta$-delayed $\gamma$ rays de-exciting the populated states in $^{154}$Gd were detected with the
VENTURE array~\cite{venture}, consisting of eight fast CeBr$_3$ scintillator detectors for timing
measurements, coupled with two Compton-suppressed Clover HPGe detectors to provide high-resolution
$\gamma$-ray energy information and to facilitate clean coincidence selection. Standard NIM electronics
were used for pulse processing, and event-by-event data were acquired through a VME-based system using the
LAMPS software~\cite{lamps}. Lifetimes  were extracted using the Generalized Centroid Difference (GCD) method~\cite{regis2016}, which relies on centroid shifts between delayed and
anti-delayed time distributions obtained from $\gamma$--$\gamma$ cascades.

\section{Analysis Procedure \& Results}

This section describes the GCD technique focusing the procedure adopted for the determination of level lifetime in the order of few tens of picoseconds and the lifetime results obtained for the 1$^-$, 1414~keV and 2$^-$, 1398~keV states in $^{154}$Gd. 

\subsection{Lifetime of 4$^+$ state of ground state band}

We start with the confirmation of experimentally available lifetime of the first-excited 4$^+$ state in $^{154}$Gd displayed in Fig.~\ref{fig1}. Details on lifetime measurement technique using VENTURE array employing GCD method can also be found in Ref.~\cite{shefali_prc}. 
\begin{figure}[ht!]
    \begin{center}
\includegraphics[width=\columnwidth]{248_347_all.jpg}
          \end{center}
\caption{The measurement has been shown with real time data for the 4$^+$ level in $^{154}$Gd. The CeBr$_3$ energy-gated projections of CeBr$_3$ (blue,solid) and Clover (red, dash-dot-dot) detectors are shown in panel (a) for decay and in panel (b) for feeder $\gamma$ rays corresponding to a cascade. The delayed (red, dash-dot-dot) and anti delayed (blue, solid) time difference spectra are also shown in panel (c). Background correction analysis are displayed in panel (d) and (e). The true centroid differences ($\Delta C_{FEP}$ = 300(7) ) due to photopeak-photopeak coincidences were obtained through determination of (1) the experimental centroid differences (including photopeak and Compton background) ($\Delta C_{expt}$= 336(6)) and (2) the correction (t$_{corr}$ = -36(4)) determined from the centroids corresponding to photopeak-Compton and Compton-Compton coincidences ($\Delta C_{BG}$). Lifetimes ($\tau$) were measured using the relation of $\Delta C_{FEP} - PRD = 2\tau$. PRD = 116(12)represents the prompt response distribution of the setup for the cascade as determined frrom Fig.~\ref{prd}. Lifetime analysis for $4^+$ level with 347 - 248 keV cascade results halflife of 65(5)ps. } 
        \label{fig1}      
    \end{figure}

\begin{figure}[ht!]
\begin{center}\includegraphics[width=\columnwidth]{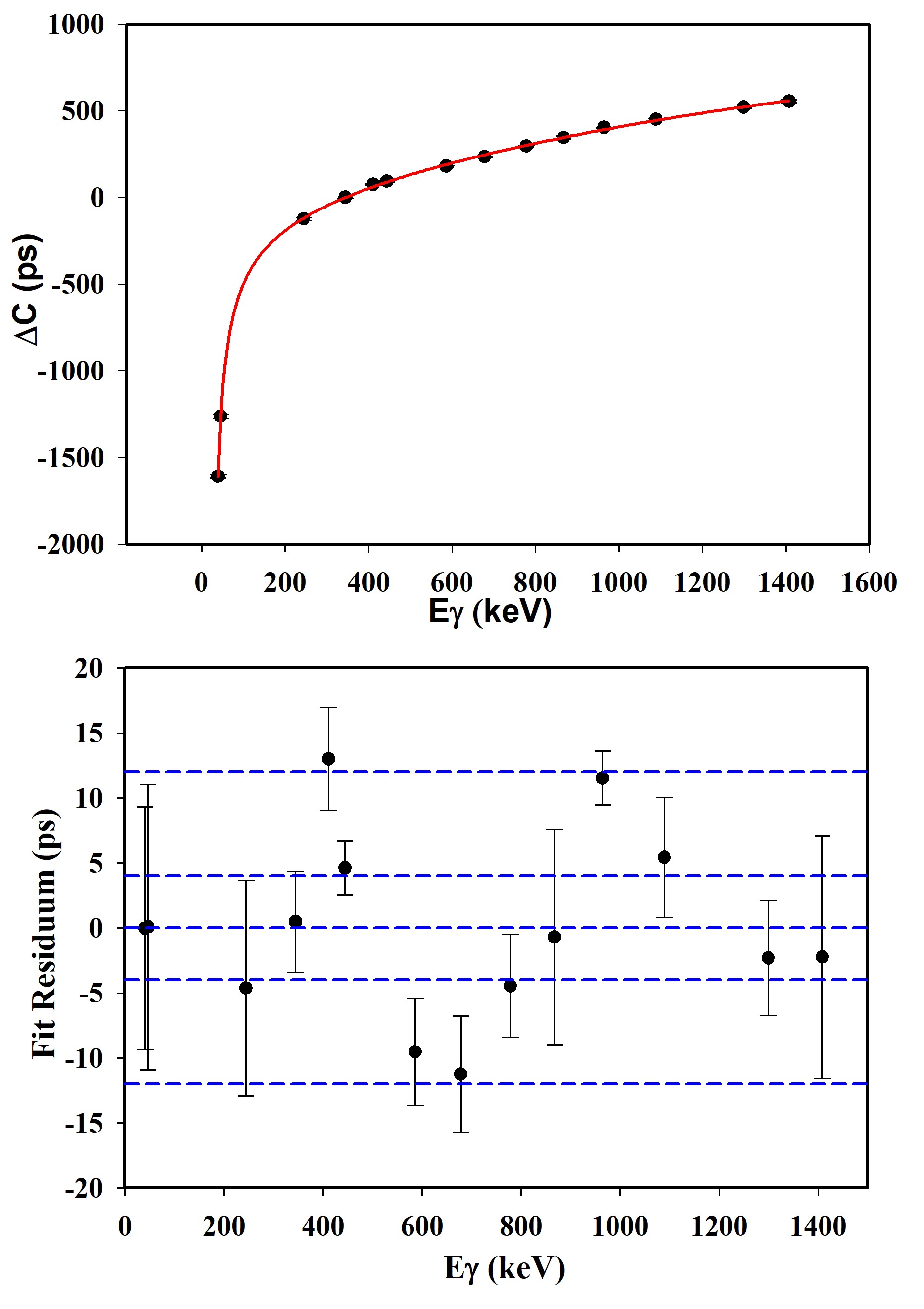}
\caption{(Color online) The PRD values, obtained with the relative time difference analysis of several cascades, are shown with blak circles as a function of energy in the upper panel. The data points have been fitted with the relation $\mathrm{PRD}(E_{\gamma}) =
\frac{a}{\sqrt{E_{\gamma} + b}}
+ c{E_{\gamma}^2} + d E_{\gamma} + e$ and the fit residuum has been attached showing a 3$\sigma$ value of 12~ps.}
\label{prd}
\end{center}
\end{figure}

The centroids of time difference ($\Delta$T) distributions, obtained with the $\gamma-\gamma$ cascade ($\gamma _{feeding} - \gamma _{decay}$) connecting the levels of interest, are measured in the GCD method. The difference between the two centroids ($\Delta C_{expt}$) of (i) delayed ($\Delta$T = T$_{\gamma _{decay}}$ - T$_{\gamma _{feeding}}$) and (ii) anti-delayed ($\Delta$T = T$_{\gamma _{feeding}}$ - T$_{\gamma _{decay}}$) time difference distributions, generated with gating the photopeaks (containing full energy and Compton events) of feeding and decaying $\gamma$ rays, are used for lifetime determination. The energy gates are shown in Fig.~\ref{fig1}(a)~and~(b) whereas the delayed and anti-delayed time distributions are shown in Fig.~\ref{fig1}(c). Narrow energy gates $\sim$ FWHM of the photopeak is selected for generating the time difference distributions and care is taken to avoid contamination from any close lying $\gamma$ transition. The measurement involves a correction for the Compton contributions (t$_{corr}$) underlying the photopeaks on which energy gates have been placed. The corrected centroid difference ($\Delta C_{FEP}$), so obtained, represents the centroid difference corresponding the full energy peak (FEP) of the $\gamma$ rays in the cascade and is related to the lifetime ($\tau$) of the intermediate state (2$\tau$  = $\Delta C _{FEP}$ - PRD).  PRD represents the Prompt Response Distribution (PRD) that is the difference of the inherent time walks of the setup at the energies of the cascade used for the measurement. The PRD is determined with reference to the 344~keV and with the cascades of $^{152}$Eu for which the intermediate levels have known lifetimes. The PRD values obtained for the present setup at different $\gamma$ energes with respect to the 344~keV transition is shown in Fig.~\ref{prd} along with the fitting of these data points to obtain the energy dependent PRD curve. The Fit residuum for PRD determination is also shown which gives rise to a standard deviation of 4~ps that has been used to determine the error in PRD values ($\delta PRD$ = 3$\sigma$ = 12~ps).
The sets of equations representing this procedure are given below while Fig.~\ref{fig1} describes the measurement technique using the first excited 4$^+$ state in $^{154}$Gd for which the lifetime value is known from literature.

\begin{align}
2\tau &= \Delta C_{\mathrm{FEP}} - \mathrm{PRD} \\
\Delta C_{\mathrm{FEP}} &= \Delta C_{\mathrm{expt}} + t_{\mathrm{corr}} \\
t_{\mathrm{corr}} &=
\frac{
\left[p/b(E_{\mathrm{decay}})\right]\, t_{\mathrm{corr}}^{(\mathrm{feeder})}
+
\left[p/b(E_{\mathrm{feeder}})\right]\, t_{\mathrm{corr}}^{(\mathrm{decay})}
}{
p/b(E_{\mathrm{feeder}}) + p/b(E_{\mathrm{decay}})
} \\
t_{\mathrm{corr}}^{(\mathrm{feeder})} &=
\frac{ C_{\mathrm{expt}} - C_{\mathrm{BG}}^{(\mathrm{feeder})} }
     { p/b_{\mathrm{feeder}} } \\
t_{\mathrm{corr}}^{(\mathrm{decay})} &=
\frac{ C_{\mathrm{expt}} - C_{\mathrm{BG}}^{(\mathrm{decay})} }
     { p/b_{\mathrm{decay}} } \\
\delta\tau &= \frac{1}{2}
\sqrt{
(\delta \Delta C_{\mathrm{expt}})^2
+
(\delta t_{\mathrm{corr}})^2
+
(\delta \mathrm{PRD})^2
} \label{gcd}
\end{align}

In the above set of equations, $\Delta C_{BG}$ values are centroid differences corresponding to the coincidences between the photopeak (of feeder or decay $\gamma$) and the Compton (of decay or feeder $\gamma$) events.  These values are obtained by interpolating the similar centroid differences measured with the particular photopeak (decay or feeder) and a range of different energy Compton (feeder or decay) events, as shown in Fig.~\ref{fig1}(d)~and~(e). The $\Delta C_{BG}$ values scaled for the peak to background ratios ({\it p/b}) are used for determining the t$_{corr}^{(feeder)}$ and t$_{corr}^{(decay)}$ giving rise to the final background correction. The error ($\delta \tau$) in the lifetime $\tau$ is determined from the errors $\delta PRD$, $\delta \Delta C_{expt}$ and $\delta t_{corr}$, which are the errors in PRD, $\Delta C_{expt}$ and t$_{corr}$, respectively.

It is observed that the lifetime $\tau$ of the 4$^+$ state measured in the present work ($t_{1/2}$ = 65(5)~ps) is well within the known values in literature (t$_{1/2}$ = 39(5)~ps from $\beta -ce$ coincidence to 68 ps from double Coulex)~\cite{ensdf}. The present result matches well with 61(4)~ps, obtained from earlier $\gamma - \gamma$ delayed coincidence work~\cite{ensdf}. 

\subsection{Lifetimes of 1$^-$, 1414~keV and 2$^-$, 1398~keV states}

\begin{figure}[ht!]
\centering
\includegraphics[width=\linewidth]{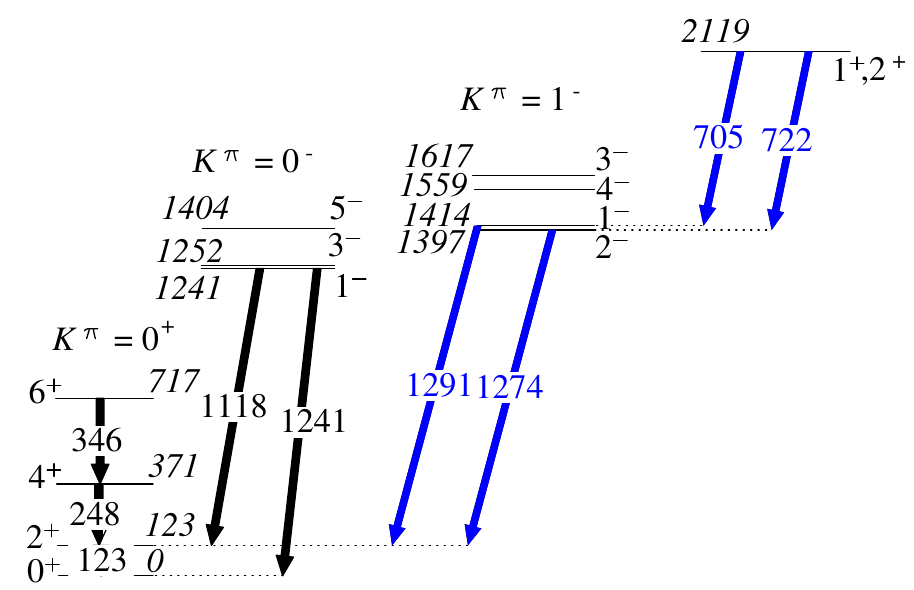}
\caption{\label{fig:level_scheme}
Partial level scheme of $^{154}$Gd showing the relevant negative-parity states and transitions used in the present work. The level scheme was
drawn following Refs.~\cite{kulp2003,kulp2004,sousa1975}. The transitions highlighted in blue were employed for the lifetime analysis.}
\end{figure}

\begin{figure}[ht!]
    \begin{center}
           \includegraphics[width=\columnwidth]{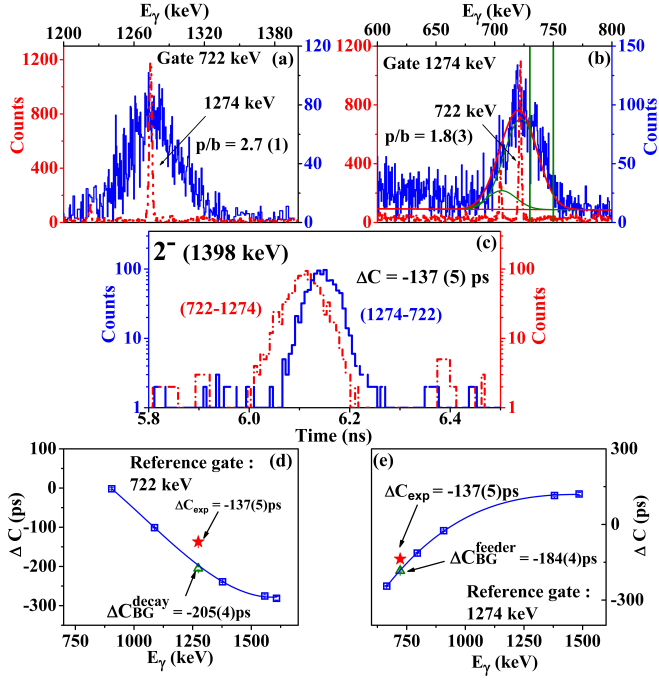}
\caption{The lifetime measurement of the 2$^-$ state at 1398~keV is displayed. The CeBr$_3$ energy-gated projections of CeBr$_3$ (blue,solid) and Clover (red, dash-dot-dot) detectors are shown in panel (a) for decay and in panel (b) for feeder $\gamma$ rays corresponding to a cascade. The delayed (red, dash-dot-dot) and anti delayed (blue, solid) time difference spectra are also shown in panel (c). Background correction analysis are displayed in panel (d) and (e). The true centroid differences ($\Delta C_{FEP}$ = -112(6) ) due to photopeak-photopeak coincidences were obtained through determination of (1) the experimental centroid differences (including photopeak and Compton background) ($\Delta C_{expt}$=-137(5)) and (2) the correction (t$_{corr}$ = 26(3) ) determined from the centroids corresponding to photopeak-Compton and Compton-Compton coincidences ($\Delta C_{BG}$). Lifetimes ($\tau$) were measured using the relation of $\Delta C_{FEP} - PRD = 2\tau$. PRD = -243(12) represents the prompt response distribution of the setup as determined from Fig.~\ref{prd}. Lifetime analysis for $2^-$ level with 722-1274 keV cascade results to $t_{1/2}$ =46(5)ps.}
        \label{fig2}
        \end{center}
    \end{figure}

\begin{figure}[ht!]
    \begin{center}
           \includegraphics[width=\columnwidth]{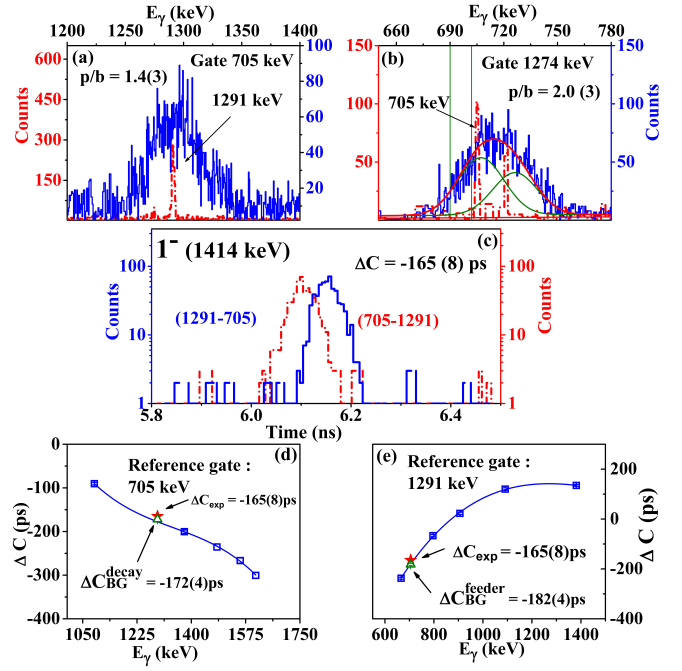}
\caption{The lifetime measurement of the 1$^-$ state at 1414~keV is displayed. The CeBr$_3$ energy-gated projections of CeBr$_3$ (blue,solid) and Clover (red, dash-dot-dot) detectors are shown in panel (a) for decay and in panel (b) for feeder $\gamma$ rays corresponding to a cascade. The delayed (red, dash-dot-dot) and anti delayed (blue, solid) time difference spectra are also shown in panel (c). Background correction analysis are displayed in panel (d) and (e). The true centroid differences ($\Delta C_{FEP}$ =-159(9) ) due to photopeak-photopeak coincidences were obtained through determination of (1) the experimental centroid differences (including photopeak and Compton background) ($\Delta C_{expt}$= -165(8)) and (2) the correction (t$_{corr}$ = 6(4)) determined from the centroids corresponding to photopeak-Compton and Compton-Compton coincidences ($\Delta C_{BG}$). Lifetimes ($\tau$) were measured using the relation of $\Delta C_{FEP} - PRD = 2\tau$. PRD = -258(12) represents the prompt response distribution of the setup as determined from Fig.~\ref{prd}. Lifetime analysis for $1^-$ level with 705-1291 keV cascade results to $t_{1/2}$ = 35(5)ps.}
        \label{fig3}
        \end{center}
    \end{figure}

After successfully reproducing the lifetime of the 4$^+$ level, the determination of the lifetimes of the negative parity states was attempted using the same procedure. The measurement for the lifetime of the first 2$^-$ and second 1$^-$ states in $^{154}$Gd are  displayed in Fig.~\ref{fig2} and Fig.~\ref{fig3}.  The relevant levels and the  $\gamma$-ray transitions involved in the present analysis are shown in Fig.~\ref{fig:level_scheme}. The lifetime of the said 2$^-$ state was measured using the 722-1274~keV cascade. However, unlike the gated projections used for the $4^+$ state (and for the 1274~keV transition), the 722~keV transition in the present case is affected by a nearby $\gamma$ ray at 705~keV, which is in coincidence with the 1291~keV transition associated with the 1274~keV line. To suppress the contribution from the 705--1291~keV cascade in the time-difference spectrum of the 722--1274~keV cascade, the 722~keV gate was carefully selected, as indicated by the two green vertical solid lines in Fig.~\ref{fig2}(b). Similarly, for the lifetime measurement of 1$^-$ state at the excitation energy $E_x$ = 1414~keV, using 705-1291~keV cascade, the 722~keV comes as the contaminant transition in the 1296~keV gate and the choice of 705~keV in the 1291~keV gate is shown in Fig.~\ref{fig3}(b). This procedure has been also discussed in detail in our earlier work~\cite{shefali_prc}. Rest of the analysis has been same as described for the 4$^+$ state.  

The measured lifetime $t_{1/2}$ of $2^-$ state comes out to be 46(5)~ps providing the first experimental data for the lowest 2$^-$ level in $^{154}$Gd. In fact, lifetime values for the 2$^-$ states are scarce in this region, specifically in Gd isotopes. For $^{156,158}$Gd and $^{160}$Gd, only lower limits on the lifetime of the $2^-$ state have been reported in the literature.  The halflife of the 1$^-$ state was found to be 35(5)~ps in the present work, providing the first reported lifetime data for $^{154}$Gd.
\begin{table}
\begin{center}
\caption{Liftimes of the $K^{\pi} = 1^-$ band in Gd isotopes. The techniques of lifetime measurements involving $\gamma-\gamma$ fast timing (FT), Doppler Broadening (DB) or Doppler Shift Attenuation Method (DSAM) are indicated.}
\begin{tabular}{ccccc}
\hline
\hline
Nucleus&E$_x$  J$^{\pi}_K$&Half life&Method&Ref.\\
&&(ps)&&\\
\hline
$^{154}$Gd&1398, 2$_1^-$&46(5)&FT&Pres. work\\
&1414, 1$_1^-$&35(5)&FT&Pres. work\\
$^{156}$Gd&1319, 2$_1^-$&$ > 3.9$&DB&\cite{Klora,156}\\ 
&1242, 1$_1^-$&$0.031^{+0.022}_{-0.009}$&DB&\cite{Klora,156}\\
$^{158}$Gd&1024, 2$_1^-$&$ > 3.5 $&DB&\cite{borner,158}\\
&917, 1$_1^-$&$1.43^{+0.19}_{-0.80}$&DB&\cite{borner,158}\\
$^{160}$Gd&1376, 2$_1^-$&$>0.389$&DSAM&\cite{lesh17,160}\\
&1351, 1$_1^-$&0.127(14)&DSAM&\cite{lesh17,160}\\
\hline
\end{tabular}
\label{tab_life}
\end{center}
\end{table}

$B(E1)$ transition rates for the de-excitation of the $2^-$ level at 1398~keV have been determined from the measured lifetime using the known branching and mixing ratios. The decay for the 2$^- \rightarrow 2^+(gsb)$ transition in $^{154}$Gd is known to be $E1+M2$ with a 96\% branching and a mixing ratio $\delta = +0.035(9)$~\cite{ensdf}. The corresponding $B(E1)$ and $B(M2)$ transition rates are determined to be 2.4$\times 10^{-6}$ W.u. and 8.4$\times 10^{-3}$ W.u., respectively. The $B(E1)$ value is comparable to the neighboring $N$ = 92 - 96 even Gd isotopes which are known to be $ < 3.2 \times 10^{-5}$ W.u. ($N$ = 92),  $ < 7.8 \times 10^{-5}$ W.u. ($N$ = 94) and $ < 2.7 \times 10^{-4}$ W.u. ($N$ = 96), indicating very similar structure for the 2$_1^-$ state in the $N$ = 90 and 92 Gd isotopes. 
The $B(E1)$ transition rates for the  1$^- \rightarrow 0^+_1$ and 1$^- \rightarrow 2^+_1$ decays have been calculated to be 5.1$\times 10^{-7}$~W.u. and 2.4$\times 10^{-6}$~W.u., respectively. These are similar to $B(E1)$ values found for the  2$^- \rightarrow 2^+_1$ decay in all the Gd isotopes where lifetime of the first 2$^-$ state is known. 

It is interesting to note that the $B(E1)$ value reported for the $1^-$ state at $E_x = 1241$~keV is much larger, $B(E1;1^- \rightarrow 0^+) = 4.1 \times 10^{-2}$~W.u., which is about four orders of magnitude higher than the $E1$ strength observed for the $2^-$ state at $E_x = 1398$~keV. This large difference in dipole strength suggests that these states are unlikely to belong to the same $K$ band. Also, it is found that the $E1$ strength for the decay of $1^-$ state at $E_x = 1414$~keV is similar to that of $2^-$ state at $E_x = 1398$~keV suggesting these two states as the candidates of the $K^{\pi} = 1^-$ band. Accordingly, the $K^{\pi} = 0^-$ band consists of the yrast odd-spin negative-parity levels in $^{154}$Gd ($1^-$, 1241~keV, 3$^-$, 1251~keV.). {The $K^\pi$ assignments used throughout this work follow, therefore, the convention adopted in earlier studies~\cite{meyer,vogel,casten1993} and the decay evaluation of ENSDF~\cite{kulp2003,kulp2004,sousa1975}. The states are represented as $J_K^{\pi}$ (referring to the spin $J$, parity $\pi$ and quantum number $K$) from here onwards, viz., the $1^-_0$, $1^-_1$ and $2^-_1$ states refer to the levels at $E_x = 1241$~keV and $E_x = 1414$~keV and $E_x = 1398$~keV, respectively. The measured halflives in $K^{\pi} = 1^-$ bands in $N$ = 90 Gd and those known in the neighboring Gd isotopes are tabulated in Table~\ref{tab_life}.

\begin{figure}[ht!]
\centering
\includegraphics[width=\linewidth]{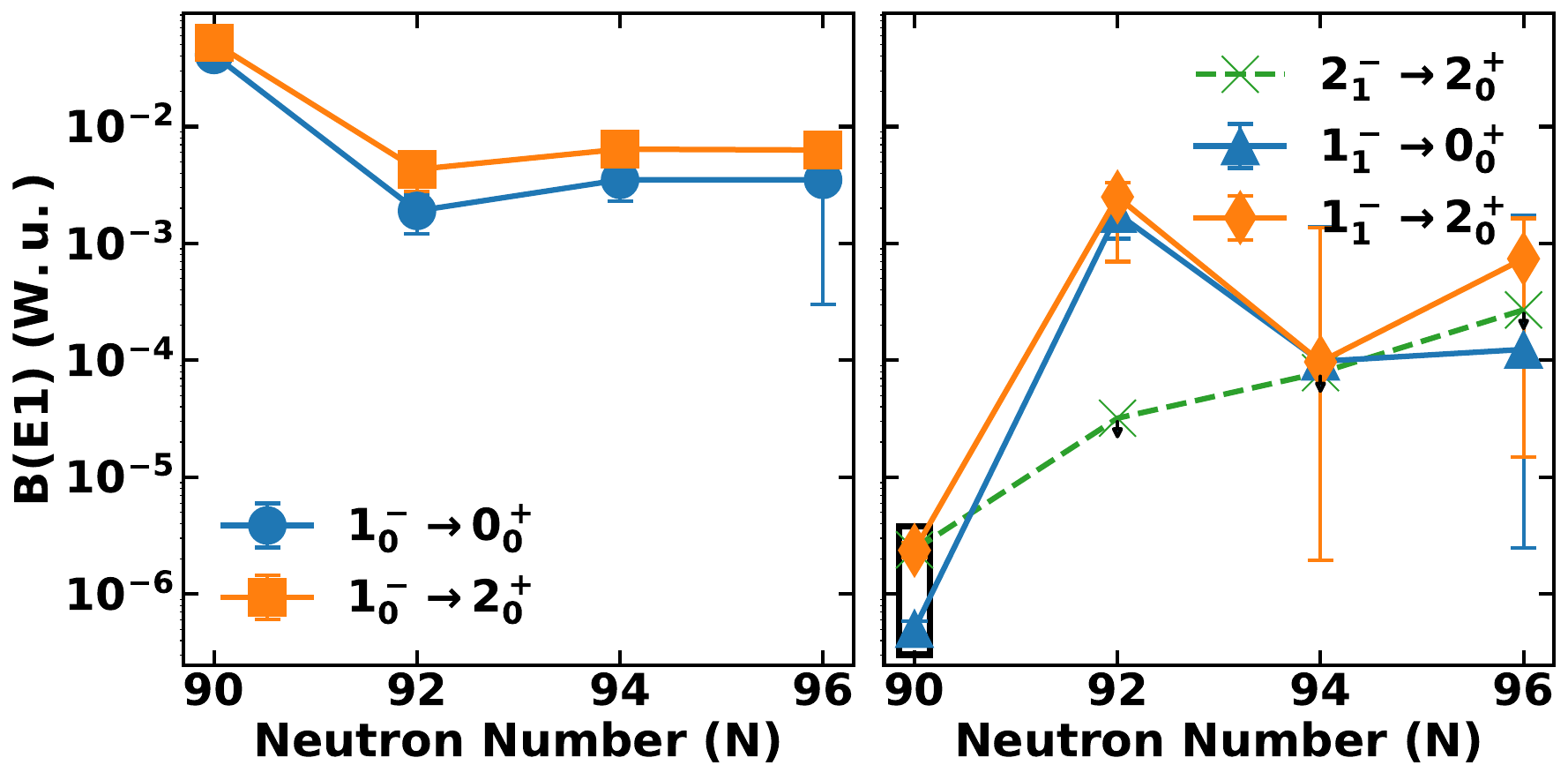}
\caption{\label{fig:BE1_systematics}
Systematics of experimental $B(E1)$ strengths in the Gd isotopic chain. Left panel: $\Delta K=0$ transitions from the $K^\pi=0^-$ band.
Right panel: $\Delta K=1$ transitions from the $K^\pi=1^-$ band. The boxed data points correspond to the present lifetime measurements in $^{154}$Gd ($N=90$), while the remaining values are adopted from evaluated data in the NNDC database~\cite{nndc}.}
\end{figure}

Figure~\ref{fig:BE1_systematics} summarizes the available experimental $B(E1)$ strengths in the Gd isotopic chain. The $\Delta K=0$ transitions from the $K^\pi=0^-$ band are generally stronger than the corresponding $\Delta K=1$ transitions from the $K^\pi=1^-$ band. The present values for the 1414-keV $1^-_1$ and 1398-keV $2^-_1$ states in $^{154}$Gd fall in the weak-strength region of the right panel, supporting their association with the weakly decaying $K^\pi=1^-$ sequence. A similar suppression of $\Delta K=1$ strength has also been reported in $^{152}$Sm at $N=90$, where the corresponding ratio is of the order of $10^{-4}$~\cite{garret2009}. This suggests that the hindrance of $\Delta K=1$ dipole transitions may be a common feature in this mass region, with $^{154}$Gd representing a more extreme case. A quantitative discussion of this suppression and its implication for band mixing is provided in the following section (Sec.~\ref{sec:band_mixing}) along with the theoretical interpretation from Gogny-HFB-based $sdf$-IBM calculation (Sec.~\ref{sec:ibm}).

\section{Theoretical Interpretation}
\label{sec:theory}

In the following subsections, the experimentally observed $B(E1)$ rates for the decay of different levels of $K^{\pi}$ bands in Gd isotopes have been interpreted in the light of the IBM calculation. A band mixing  calculation has been performed to understand the role of configuration mixing in the suppression of B(E1) rates from different K configurations. 

In Sec.~\ref{sec:ibm}, we discuss the Gogny-HFB-based $sdf$-IBM calculations for the low-lying negative-parity states in the even-even Gd isotopes and compare the calculated excitation energies and $E1$ transition strengths with the available experimental data. In Sec.~\ref{sec:band_mixing}, we present a band-mixing analysis for $^{154}$Gd to examine the observed hindrance of the $\Delta K=1$ dipole transitions.

\subsection{Gogny-HFB-based $sdf$-IBM calculation}
\label{sec:ibm}
 We perform calculations for the even-even $^{148-158}$Gd isotopes using the IBM framework mapped from a mean-field potential. The starting point is axially-symmetric quadrupole and octupole deformation moments-constrained Hartree-Fock-Bogoliubov (HFB) calculations with Gogny force to obtain the corresponding mean-field energy surfaces and HFB states for these Gd isotopes. The quadrupole and octupole deformation moments are simplified into the standard $\beta_2$ and $\beta_3$ deformation parameters. These mean-field energy surfaces are then mapped into the IBM energy surfaces using the procedure discussed in Ref.~\cite{nomura2015}. Such mapping allows us to fix the IBM Hamiltonian parameters for calculating energies and $E1$ transition probabilities of low-lying nuclear states. To describe the parity-changing $E1$ transitions, the IBM framework is chosen to include both positive ($s$ and $d$) and negative ($f$) parity bosons. The $sdf$-Hamiltonian used is given by 
\begin{equation}
    \hat{H}=\epsilon_d \hat{n}_d + \epsilon_f \hat{n}_f + \kappa_2 \hat{Q}_2.\hat{Q}_2 + \kappa'_2 \hat{L}_d.\hat{L}_d + \kappa_3 \hat{Q}_3.\hat{Q}_3
\end{equation}
where the first and second terms denote the number operator with $\epsilon_d$ and $\epsilon_f$ relative single-particle energies for $d$ and $f$ bosons with respect to the $s$ boson. The third and fifth term represent the quadrupole-quadrupole and octupole-octupole interaction with respective strengths $\kappa_2$ and $\kappa_3$. The fourth term is the rotational term relevant for the $sd$ bosonic space. The quadrupole operator $\hat{Q}_2$ is defined as 
\begin{equation}
    \hat{Q}_2 = s^\dagger \tilde{d} + d^\dagger \tilde{s} + \chi_{dd} [d^\dagger \times \tilde{d}]^{(2)} + \chi_{ff} [f^\dagger \times \tilde{f}]^{(2)}
\end{equation}
with parameters $\chi_{dd}$ and $\chi_{ff}$. The octupole operator $\hat{Q_3}$ is defined by 
\begin{equation}
    \hat{Q}_3 = s^\dagger \tilde{f} + f^\dagger \tilde{s} + \chi_{df} [d^\dagger \times \tilde{f} + f^\dagger \times \tilde{d} ]^{(2)}
\end{equation}
with $\chi_{df}$ being a parameter. The angular momentum operator $\hat{L}_d$ in the fourth term can be expressed as
\begin{equation}
    \hat{L}_d = \sqrt{10} [d^\dagger \times \tilde{d}]^{(1)}
\end{equation}
The dipole-dipole interaction term $\hat{L_d}.\hat{L_f}$ with $\hat{L_f}=\sqrt{28}[d^\dagger \times \tilde{f}]^{(1)}$ is neglected as shown to have little relevance for low-energy states~\cite{cottle}. 

The framework inclusive of quadrupole and octupole degrees of freedom can hence be used to study the impact of octupole correlations on the low-lying collective spectra of Gd isotopes, a mass region where such correlations play an important role. The calculations utilizing this framework for $^{146-156}$Sm and $^{148-158}$Gd isotopes were already successfully shown earlier~\cite{nomura2015}. The electric dipole ($E1$) and electric octupole ($E3$) transitions respectively from the first $1^-$ and $3^-$ states to the ground states were also explained well. In this work, we extend and examine these calculations for the newly measured $E1$ transitions connecting the low-lying negative-parity states to the $2^+$ state of the ground state band, especially in $^{154}$Gd.

From the diagonalization of the $sdf$-IBM Hamiltonian treating positive- and negative-parity bosons on an equal footing, we obtain energies, total spin and parity of the states. These calculated values are compared with available experimental data in Fig.~\ref{fig:en}
for the low-lying negative-parity states in even-even $^{148-158}$Gd isotopes. As the calculations cannot provide the firm $K$-assignment, we have considered the lowest energy $K$ sequence as $K^{\pi} = 0^-$ and the next one as the $K^{\pi} = 1^-$, following the experimental observation in $^{154}$Gd and similarly, for the other Gd isotopes. 
\begin{figure}[ht!]
\centering
\includegraphics[width=\linewidth]{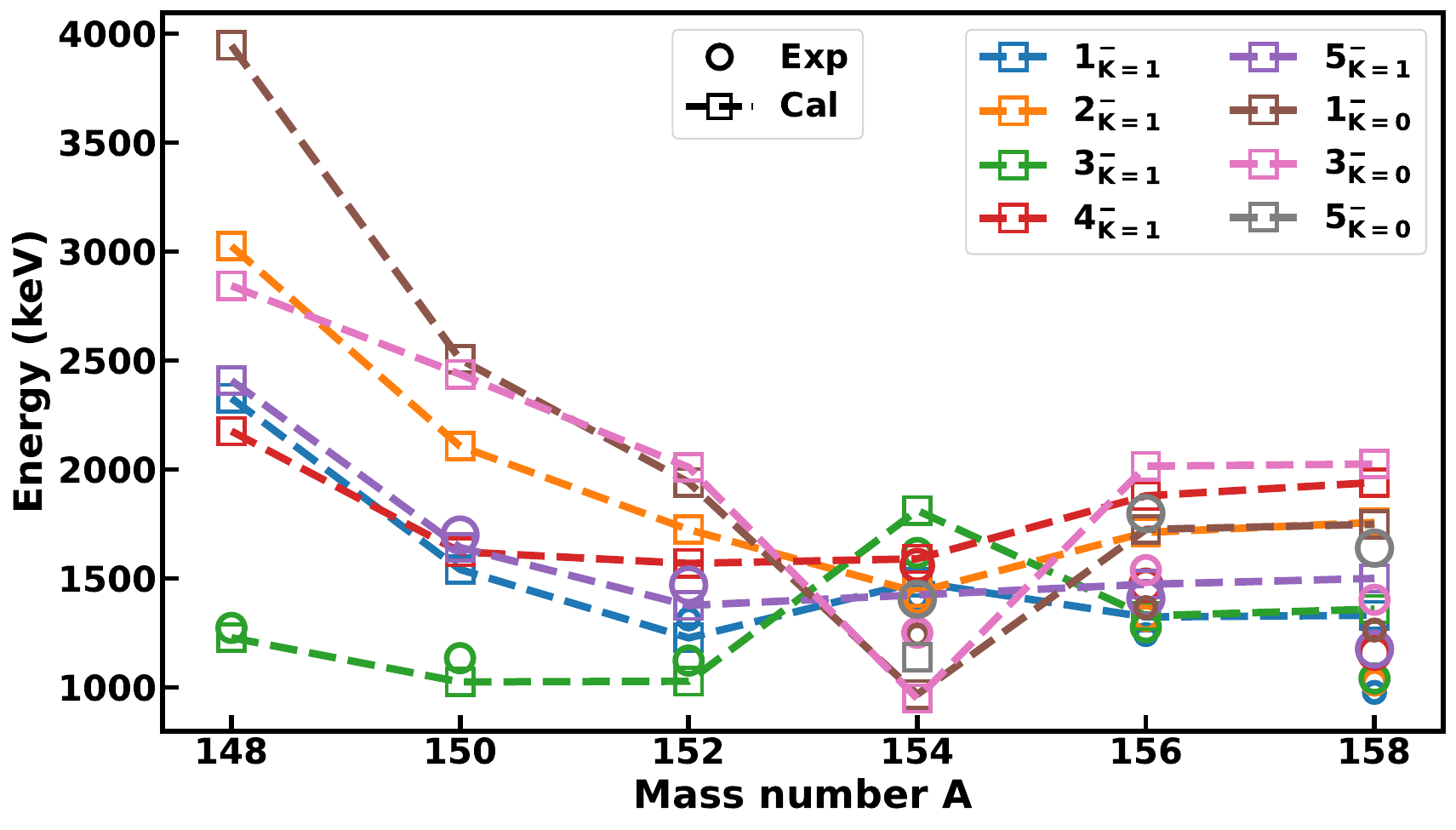}
\caption{\label{fig:en}  Experimental and calculated energies of the low-lying states in even-even $^{148-158}$Gd isotopes. The states are labeled with $J^\pi_K$ referring to the spin $J$, parity $\pi$ and quantum number $K$.}
\end{figure}
The calculations reproduce the overall trend of the experimental excitation energies reasonably well across the Gd isotopic chain. In particular, the calculated energies of the $1^-_1$ and $2^-_1$ states in $^{154}$Gd, which are the focus of the present work, are in good agreement with the experimental values. The calculations also reproduce the unusual ordering in which the $2^-_1$ state lies
slightly below the $1^-_1$ state in $^{154}$Gd. However, for the heavier isotopes $^{156}$Gd and $^{158}$Gd the calculations tend to overestimate the excitation energies of these states, indicating a gradual deviation beyond $N=90$.
For the lighter isotopes $^{148,150,152}$Gd, the calculated
energies of the $3^-_1$ states reproduce the available experimental data reasonably well. Experimental information on the $1^-_1$ state is not available for $^{148}$Gd and $^{150}$Gd, but the calculated energy in $^{152}$Gd agrees well with the measured value. In the isotopes $^{154,156,158}$Gd the $1^-_1$ and $3^-_1$ states
are found to lie close in energy both experimentally and in the calculations. The calculated energies also follow the experimental systematics of the $5^-_1$ states across the $^{148-158}$Gd isotopes, although small discrepancies are observed in $^{154}$Gd and $^{158}$Gd.

Figure~\ref{fig:en2} shows the variation of the low-lying energy levels of the $K^{\pi}=0^-$ and $K^{\pi}=1^-$ bands as a function of neutron number in the Gd isotopes. 
The data show that the excitation energies of the $K^{\pi}=0^-$ band exhibit a minimum around $N=90$, whereas those of the $K^{\pi}=1^-$ sequence display a maximum near the same neutron number. This opposite
trend reflects a transitional behaviour in the relative ordering of the two octupole structures. The lower panel illustrates the energy difference $\Delta E = E(1^-_{1}) - E(1^-_{0})$, which highlights the evolution of the relative band ordering across the isotopic chain. The change in
the sign of $\Delta E$ reflects the inversion between the $K^{\pi}=0^-$ and $K^{\pi}=1^-$ configurations, a feature previously discussed by Casten \textit{et al.}~\cite{casten1993}.
\begin{figure}[ht!]
\centering
\includegraphics[width=\linewidth]{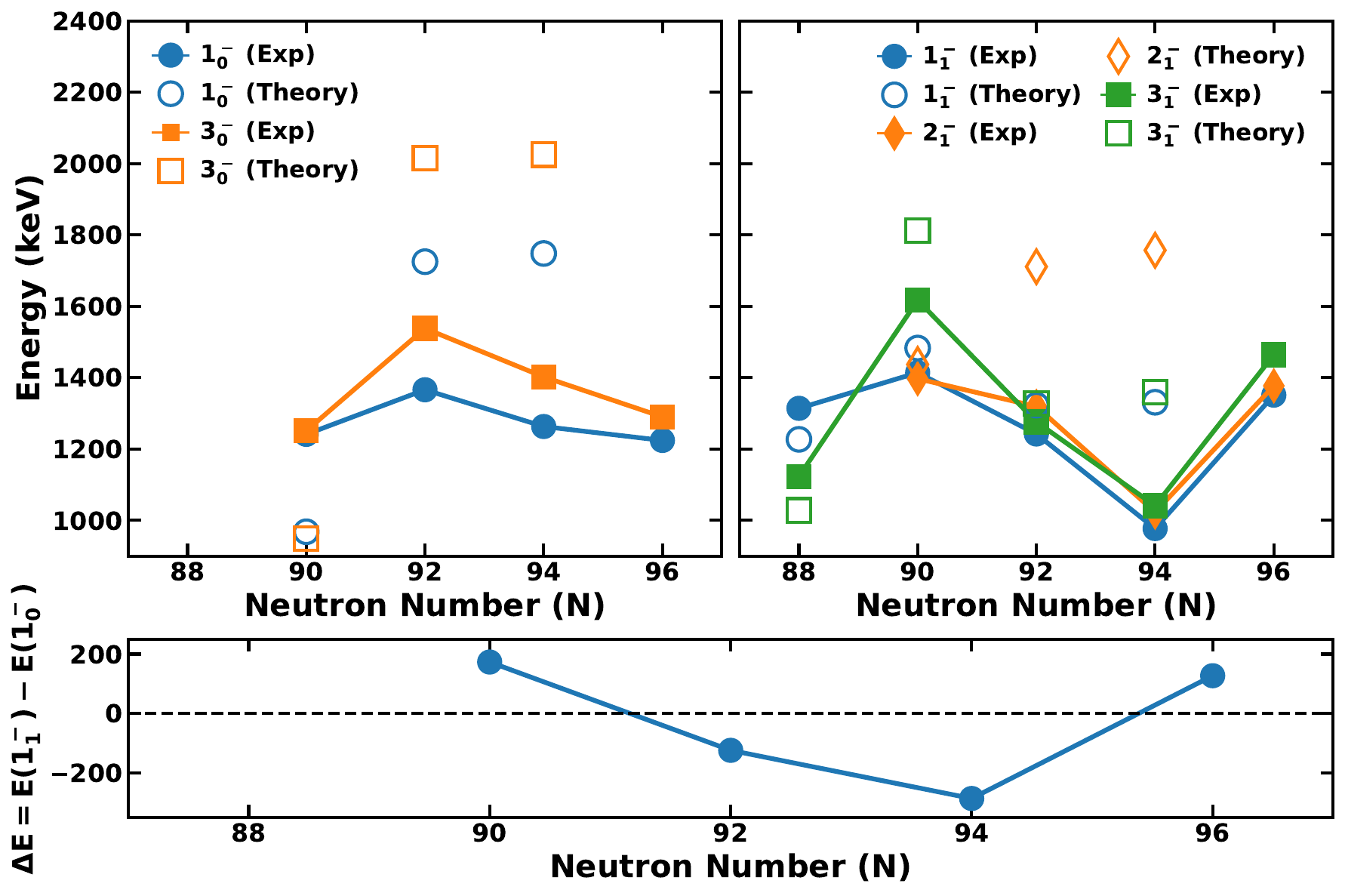}
\caption{\label{fig:energy_exp}  Systematics of low-lying negative-parity levels in the Gd isotopic chain as a function of neutron number.
Upper left panel: experimental and calculated excitation energies of the $K^\pi=0^-$ band ($1^-$ and $3^-$ states).
Upper right panel: corresponding energies for the $K^\pi=1^-$ band ($1^-$, $2^-$, and $3^-$ states).
Lower panel: energy difference $\Delta E = E(1^-_{1}) - E(1^-_{0})$ illustrating the change in band sequence.
The sign change of $\Delta E$ indicates the inversion between the $K=0^-$ and $K=1^-$ configurations across the isotopic chain.}
\label{fig:en2}
\end{figure}
\begin{table*}[ht!]
\centering
\caption{Calculated and experimental (where available) $B(E1)$ values (in Weisskopf units) for selected transitions in $^{148-158}$Gd. 
The subscripts in the (spin) states indicate the $K$ assignments. Experimental data, other than those measured in the present work, 
are taken from ~\cite{ensdf,156,158,160}. The values obtained in the present work are shown in \textbf{bold}.}
\renewcommand{\arraystretch}{1.15}
\begin{tabular}{c| cc|cc|cc}
\hline\hline
$N$ & \multicolumn{6}{c}{$B(E1)$ (W.u.)} \\
\cline{2-7}

& \multicolumn{2}{c}{$1^-_{0} \rightarrow 0^+$}
& \multicolumn{2}{c}{$1^-_{1} \rightarrow 0^+$}
& \multicolumn{2}{c}{$2^-_{1} \rightarrow 2^+$}\\

\cline{2-7}
& Calc. & Expt. & Calc. & Expt. & Calc. & Expt. \\
\hline

84 & $3.1\times10^{-5}$ & --
   & $9.1\times10^{-5}$ & --
   & $2.4\times10^{-4}$ & -- \\

86 & $5.4\times10^{-5}$ & --
   & $3.7\times10^{-4}$ & --
   & $5.2\times10^{-5}$ & -- \\

88 & $2.0\times10^{-4}$ & --
   & $1.1\times10^{-3}$ & --
   & $8.2\times10^{-4}$ & -- \\

90 & $2.4\times10^{-3}$ & $4.1\times10^{-2}$
   & $6.7\times10^{-4}$ & $\mathbf{(5.1^{+0.8}_{-0.6})\times10^{-7}}$
   & $2.4\times10^{-3}$ & $\mathbf{(2.4^{+0.4}_{-0.3})\times10^{-6}}$ \\

92 & $1.0\times10^{-3}$ & $(1.9 \pm 0.7)\times10^{-3}$
   & $3.8\times10^{-3}$ & $(1.8 \pm 0.7)\times10^{-3}$
   & $3.7\times10^{-3}$ & $ < 3.2\times10^{-5}$ \\

94 & $1.5\times10^{-3}$ & $(3.5 \pm 1.2)\times10^{-3}$
   & $4.2\times10^{-3}$ & $(9.8^{+128}_{-11})\times10^{-5}$
   & $4.5\times10^{-3}$ & $ <7.8\times10^{-5}$ \\


\hline\hline
\end{tabular}
\label{tab:E1}
\end{table*}


Electric reduced transition probabilities are calculated by 
\begin{equation}
    B(EL; J \rightarrow J') = \frac{|\langle J' || \hat{T}(EL) || J \rangle |^2}{2J+1} 
\end{equation}
using the IBM wave functions obtained from the diagonalization, where $J$ and $J'$ denote the spins for initial and final states, respectively. In this work, we focus on electric dipole $E1$ transition probabilities defined by the operator $\hat{T}^{(E1)} = e_1 [d^\dagger \times \tilde{f} + f^\dagger \times \tilde{d}]^{(1)}$ with $e_1=0.01eb^{1/2}$  the electric dipole boson effective charge kept constant for all considered Gd isotopes as used in Ref.~\cite{nomura2015}. These effective charges have been found to provide reasonable agreement to the experimental data; however, they are not derived microscopically. This introduces an overall scale uncertainty in the calculated $E1$ transition probabilities, which is not adjusted here in light of the new measurement in order to assess the suitability of previously adopted values.

A comparison of the experimental and calculated $B(E1)$ values (in Weisskopf units) for three transitions is presented in Table~\ref{tab:E1} with the current measurement highlighted in bold. It is observed that the 1$_{0}^- \rightarrow 0^+$ transitions are in better agreement with experiment. The same is valid also for the 1$_{1}^{-} \rightarrow 0^+$ decay at $N$ = 92 $^{156}$Gd. However, the calculation overestimates the $B(E1;$$1_1^- \rightarrow 0^+$) rate both at $N$ = 90 and $N$ = 94 and all the $B(E1 ; 2$$_{1}^- \rightarrow 2^+$) transitions throughout the Gd chain. In order to understand the effect of effective charges used in the calculation, one may also compare the $B(E2)$ values for the intra-band and interband decays from these $K^{\pi}$ sequences. However, only the upper limits of interband $B(E2; 2$$_1^- \rightarrow 3_0^-$) $<$0.74~W.u and $B(E2; 2$$_1^- \rightarrow 1_0^-$) $<$0.78~W.u, could be derived from the present experimental data in $^{154}$Gd while no in-band transition is known.  
Similar discrepancies in the magnitude of the $E1$ strength due to the choice of effective charge were reported earlier for several $B(E1)$ transitions in $N$ = 90 $^{152}$Sm by Garrett et al.~\cite{garret2009}.

\subsection{Band-mixing calculation}
\label{sec:band_mixing}
Motivated by the strong suppression of the $\Delta K=1$ strengths shown in Fig.~\ref{fig:BE1_systematics}, we next examine whether the observed weak $E1$ decay can be understood in terms of mixing between the $K^\pi=0^-$ and $K^\pi=1^-$ bands. In the absence of band mixing, electric transitions between rotational
bands with quantum numbers $K$ and $K'$ are governed by the rotational
intensity rule~\cite{Mikhailov66}
\begin{equation}
\begin{aligned}
&\langle K'J'||\hat T(E\lambda)||KJ\rangle \\
&=
\sqrt{(2J+1)(1+\delta_{K0})}
(JK\,\lambda\,K'-K|J'K') \\
&\quad\times
\Bigl[
M^{(0)}_{KK'}
+
M^{(1)}_{KK'}
\{J'(J'+1)-J(J+1)\}
\Bigr].
\end{aligned}
\label{e_mikhailov}
\end{equation}
where $(JK\,\lambda K'-K|J'K')$ is a Clebsch-Gordan coefficient~\cite{Shalit63,Talmi93}
and $M^{(i)}_{KK'}$ is an intrinsic matrix element
for $E\lambda$ transitions from band $K$ to band $K'$.
The term in $M^{(0)}_{KK'}$ in Eq.~(\ref{e_mikhailov}) is obtained
if the intrinsic matrix element is independent
of the components $I_\pm$  of the rotational angular momentum.
The term in $M^{(1)}_{KK'}$ arises in first order
if this dependence is taken into account~\cite{BM75}.
If $M^{(1)}_{KK'}=0$, the expression~(\ref{e_mikhailov})
reduces to the Alaga rule~\cite{Alaga55}.

For the two negative-parity bands in $^{154}$Gd, the intrinsic dipole
matrix elements for the $K^\pi=0^-\rightarrow K^\pi=0^+$ and
$K^\pi=1^-\rightarrow K^\pi=0^+$ transitions are denoted by
$M^{(0)}_{00}$ and $M^{(0)}_{10}$, respectively. It is expected that
$\Delta K=1$ transitions are hindered compared with transitions that
preserve $K$, i.e.
\begin{equation}
M^{(0)}_{10}\ll M^{(0)}_{00}.
\label{eq:matrix_element_hierarchy}
\end{equation}
The ratio of the intrinsic dipole matrix elements is written as
\begin{equation}
\zeta=\frac{M^{(0)}_{10}}{M^{(0)}_{00}} .
\label{eq:zeta_def}
\end{equation}

In the Alaga limit, the branching ratios for the $1^-$ states are
\begin{equation}
\frac{B(E1;1^-_{K=0}\rightarrow0^+)}
     {B(E1;1^-_{K=0}\rightarrow2^+)}
=\frac{1}{2},
\qquad
\frac{B(E1;1^-_{K=1}\rightarrow0^+)}
     {B(E1;1^-_{K=1}\rightarrow2^+)}
=2 .
\label{eq:alaga_ratios}
\end{equation}
The corresponding experimental ratios in $^{154}$Gd are
\begin{equation}
R_{K=0}=0.79(11),
\qquad
R_{K=1}=0.21^{+0.07}_{-0.05},
\label{eq:exp_ratios}
\end{equation}
The $K=0$ ratio is reasonably close to the
Alaga expectation, whereas the $K=1$ ratio differs strongly from the
pure-$K$ value. In addition, the ratio
\begin{equation}
\frac{B(E1;1^-_{K=1}\rightarrow J^+_{K=0})}
     {B(E1;1^-_{K=0}\rightarrow J^+_{K=0})}
\sim 10^{-5}
\label{eq:be1_ratio_k1_k0}
\end{equation}
is very small, implying an unusually small value of $\zeta$.

We now consider mixing between the two $1^-$ states resulting from the
Coriolis interaction between the $K^\pi=0^-$ and $K^\pi=1^-$ bands~\cite{Kocbach1970}. If one associates the $1^-$, 1241~keV ($1^-$, 1414~keV) level with $K^\pi=0^-$ ($K^\pi=1^-$),
the corresponding physical states are represented as $|1^-_1\rangle$ and $|1^-_2\rangle$, respectively, for the following discussion, then we get 
\begin{align}
|1^-_1\rangle
&=
c_{10}|1^-_{K=0}\rangle
+
c_{11}|1^-_{K=1}\rangle ,
\nonumber\\
|1^-_2\rangle
&=
c_{11}|1^-_{K=0}\rangle
-
c_{10}|1^-_{K=1}\rangle ,
\label{eq:mixed_states}
\end{align}
with
\begin{equation}
c_{10}^{2}+c_{11}^{2}=1 .
\label{eq:mixing_norm}
\end{equation}
Introducing
\begin{equation}
\epsilon=\frac{c_{11}}{c_{10}},
\label{eq:epsilon_def}
\end{equation}
the $E1$ branching ratios can be expressed in terms of $\zeta$ and
$\epsilon$.

In lowest order, the relevant branching ratios are
\begin{align}
\frac{B(E1;1^-_1\rightarrow0^+_1)}
     {B(E1;1^-_1\rightarrow2^+_1)}
&=
2\left(
\frac{c_{11}M^{(0)}_{10}
      -c_{10}\sqrt{2}M^{(0)}_{00}}
     {c_{11}M^{(0)}_{10}
      +2c_{10}\sqrt{2}M^{(0)}_{00}}
\right)^2
\nonumber\\
&=
2\left(
\frac{\epsilon\zeta-\sqrt{2}}
     {\epsilon\zeta+2\sqrt{2}}
\right)^2 ,
\nonumber\\
\frac{B(E1;1^-_2\rightarrow0^+_1)}
     {B(E1;1^-_2\rightarrow2^+_1)}
&=
2\left(
\frac{c_{11}\sqrt{2}M^{(0)}_{00}
      +c_{10}M^{(0)}_{10}}
     {2c_{11}\sqrt{2}M^{(0)}_{00}
      -c_{10}M^{(0)}_{10}}
\right)^2
\nonumber\\
&=
2\left(
\frac{\sqrt{2}\epsilon+\zeta}
     {2\sqrt{2}\epsilon-\zeta}
\right)^2 .
\label{eq:mixing_ratios_lowest}
\end{align}

In the absence of mixing, $\epsilon=0$, one recovers the Alaga rules.
Also, if the intrinsic matrix element $M^{(0)}_{10}$ is much smaller
than $M^{(0)}_{00}$, i.e. $\zeta\rightarrow0$, both ratios approach the
Alaga value associated with the $\Delta K=0$ transition.

The ratios of $B(E1)$ values from the two negative-parity bands to the
ground-state band are modified to
\begin{align}
\frac{B(E1;1^-_2\rightarrow0^+_1)}
     {B(E1;1^-_1\rightarrow0^+_1)}
&=
\left(
\frac{-c_{11}\sqrt{2}M^{(0)}_{00}
      -c_{10}M^{(0)}_{10}}
     {-c_{10}\sqrt{2}M^{(0)}_{00}
      +c_{11}M^{(0)}_{10}}
\right)^2
\nonumber\\
&=
\left(
\frac{\sqrt{2}\epsilon+\zeta}
     {\epsilon\zeta-\sqrt{2}}
\right)^2 ,
\nonumber\\
\frac{B(E1;1^-_2\rightarrow2^+_1)}
     {B(E1;1^-_1\rightarrow2^+_1)}
&=
\left(
\frac{2c_{11}\sqrt{2}M^{(0)}_{00}
      -c_{10}M^{(0)}_{10}}
     {2c_{10}\sqrt{2}M^{(0)}_{00}
      +c_{11}M^{(0)}_{10}}
\right)^2
\nonumber\\
&=
\left(
\frac{2\sqrt{2}\epsilon-\zeta}
     {\epsilon\zeta+2\sqrt{2}}
\right)^2 .
\label{eq:be1ratio2mix0}
\end{align}
The expressions in Eq.~(\ref{eq:be1ratio2mix0}) show that small ratios
may arise from destructive interference in the numerator between the
intrinsic matrix-element ratio $\zeta$ and the mixing amplitude ratio
$\epsilon$. Nevertheless, in this approximation it is still difficult to
explain why both ratios in Eq.~(\ref{eq:be1_ratio_k1_k0}), with
$J^\pi=0^+$ and $J^\pi=2^+$, are so small, since in one numerator
$\zeta$ and $\epsilon$ have the same sign, while in the other they have
opposite signs.

The first-order dependence of the intrinsic matrix elements on the
components $I_\pm$ of the rotational angular momentum was therefore
included. Since the hierarchy
\begin{equation}
M^{(1)}_{10}\ll M^{(0)}_{10}\ll M^{(0)}_{00}
\label{eq:matrix_hierarchy}
\end{equation}
is expected, $M^{(1)}_{10}$ was neglected. The remaining correction is
described by
\begin{equation}
\eta=\frac{M^{(1)}_{00}}{M^{(0)}_{00}} .
\label{eq:eta_def}
\end{equation}
The four independent ratios involving the two $1^-$ states then become
\begin{align}
\frac{B(E1;1^-_1\rightarrow0^+)}
     {B(E1;1^-_1\rightarrow2^+)}
&=
2\left[
\frac{\epsilon\zeta-\sqrt{2}(1-2\eta)}
     {\epsilon\zeta+2\sqrt{2}(1+4\eta)}
\right]^2 ,
\nonumber\\
\frac{B(E1;1^-_2\rightarrow0^+)}
     {B(E1;1^-_2\rightarrow2^+)}
&=
2\left[
\frac{\sqrt{2}\epsilon(1-2\eta)+\zeta}
     {2\sqrt{2}\epsilon(1+4\eta)-\zeta}
\right]^2 ,
\nonumber\\
\frac{B(E1;1^-_2\rightarrow0^+)}
     {B(E1;1^-_1\rightarrow0^+)}
&=
\left[
\frac{\sqrt{2}\epsilon(1-2\eta)+\zeta}
     {\epsilon\zeta-\sqrt{2}(1-2\eta)}
\right]^2 ,
\nonumber\\
\frac{B(E1;1^-_2\rightarrow2^+)}
     {B(E1;1^-_1\rightarrow2^+)}
&=
\left[
\frac{2\sqrt{2}\epsilon(1+4\eta)-\zeta}
     {\epsilon\zeta+2\sqrt{2}(1+4\eta)}
\right]^2 .
\label{eq:mixing_ratios_firstorder}
\end{align}

The parameters $\zeta$, $\epsilon$, and $\eta$ were obtained by equating
three of the above ratios to their corresponding experimental values.
Among the possible solution branches, the one which best reproduces the
measured $B(E1;2^-_1\rightarrow2^+_1)$ value gives
\begin{equation}
\zeta = -0.0095^{+6}_{-6},
\qquad
\epsilon = 0.0028^{+3}_{-3},
\qquad
\eta = -0.037^{+11}_{-10}
\label{eq:fit_zeta_epsilon_eta}
\end{equation}

An equivalent fit can be performed directly in terms of the intrinsic
matrix elements and mixing amplitudes. With $M^{(1)}_{10}=0$ and
$c_{10}^{2}+c_{11}^{2}=1$, the preferred solution gives approximately
\begin{equation}
\begin{aligned}
M^{(0)}_{00} &\simeq -0.231, \qquad
M^{(0)}_{10} \simeq 0.00222, \\
M^{(1)}_{00} &\simeq 0.00842, \qquad
c_{11} \simeq 0.00281 .
\end{aligned}
\label{eq:preferred_matrix_solution}
\end{equation}
with $c_{10}>0$. This solution predicts
\begin{equation}
B(E1;2^-_1\rightarrow2^+_1)
\simeq 2.5\times10^{-6}~{\rm W.u.},
\label{eq:predicted_2minus_be1}
\end{equation}
in good agreement with the measured value.

The small value of the mixing amplitude,
$\epsilon \simeq 0.0028$,
indicates only a very small admixture between the two bands;
this conclusion remains unchanged within the quoted uncertainty. The small value of
$|\zeta|$ further indicates that the intrinsic $\Delta K=1$ dipole
matrix element is strongly hindered compared with the $\Delta K=0$
matrix element.

\section{Summary}


In the present work, we have reported the first lifetime measurements of the $1^-$ (1414~keV) and $2^-$ (1398~keV) in N = 90 $^{154}$Gd. Using the measured lifetimes together with the known branching ratios and multipole-mixing
information, absolute $E1$ transition strengths were deduced. The extracted $B(E1)$ values are of the order of $10^{-7}$--$10^{-6}$~W.u.,
showing that the dipole decays from these levels are strongly hindered. 

In particular, the measured $B(E1;2^-_1\rightarrow2^+_1)$ value represents, to our knowledge, the only confirmed absolute strength for this transition presently available in the Gd isotopic chain, where only upper limits are known for the heavier isotopes. The ratios of the measured B(E1) rates from the 1414~keV 1$^-$ level and the absolute B(E1) rate from the 2$_1^-$, 1398~keV level could be explained with a simple band mixing calculation considering these two levels as the candidates of $K^\pi=1^-$ configuration.

The present data, in turn, provide an important experimental anchor point for the systematics of low-lying $\Delta K=1$ dipole strength at $N=90$. The extracted $B(E1)$ values are strongly suppressed compared with the corresponding $\Delta K=0$ transitions from the $K^\pi=0^-$ band, indicating strong $\Delta K=1$ hindrance and limited $K$ mixing in $^{154}$Gd. The preferred solution from the band mixing analysis gives a very small mixing amplitude between the $K^\pi=0^-$ and $K^\pi=1^-$ bands. The same solution reproduces the measured
$B(E1;2^-_1\rightarrow2^+_1)$ strength, indicating that the observed hindrance can be understood as a consequence of weak intrinsic $\Delta K=1$ dipole strength together with limited Coriolis mixing.

Both the excitation-energy and B(E1) systematics suggest that the stronger octupole-related dipole collectivity is concentrated in the $K^\pi=0^-$ branch, whereas the $K^\pi=1^-$ sequence carries much weaker dipole strength at N = 90.

The Gogny-HFB-based $sdf$-IBM calculations reproduce the overall evolution of the negative-parity excitation energies, including the unusual $E(2^-_1)<E(1^-_1)$ ordering in $^{154}$Gd, but significantly overestimate the observed $E1$ strengths of the $K^\pi=1^-$ states. Since the electric dipole boson effective charge may also be state dependent, and no other definite $B(E1;2^-_1\rightarrow2^+)$ values are available for these Gd isotopes apart from the present measurement and the upper limits in heavier Gd isotopes, future lifetime measurements will be important for testing this possibility. The framework could also be extended by including higher-order terms in the $E1$ operator, or by explicitly incorporating the $p$ boson into the model space. Such improvements may also help in treating spurious-state contributions associated with center-of-mass corrections in electric dipole transitions. Future work will focus on these aspects in conjunction with new experimental data.


\section{Acknowledgement}

The authors gratefully acknowledge the effort of K130 cyclotron operation staff for providing stable light ion beams during experiment.  B.~Maheshwari gratefully acknowledges the funding support from ANRF (India), RJF/2025/000092, and to the HORIZON-MSCA-2023-PF-01 project, ISOON, under grant number 101150471 at GANIL (France).

\end{document}